\documentstyle[pre,aps,psfig]{revtex}
\newcommand{\nc}{\newcommand}
\newcommand{\rnc}{\renewcommand}
\rnc{\/}{\over}
\nc{\half}{{1\/2}}
\nc{\<}{\langle}
\rnc{\>}{\rangle}
\rnc{\/}{\over}
\rnc{\(}{\left (}
\rnc{\)}{\right )}
\rnc{\[}{\left [}
\rnc{\]}{\right ]}
\nc{\ket}[1]{|#1\>}
\nc{\bra}[1]{\<#1|}
\nc{\mrm}[1]{{\rm #1}}
\rnc{\Re}{\mrm{Re}\,}
\rnc{\Im}{\mrm{Im}\,}
\nc{\Tr}{\mrm{Tr}\,}
\nc{\Sgn}{\mrm{Sgn}}
\nc{\e}{\mbox{e}}
\rnc{\i}{{\rm i}}
\nc{\dos}{\<d\>}
\newcommand{\inodot}{\char'020}

\begin{document}
\draft{
\title{  Spectral correlations in systems  undergoing
a transition from periodicity to disorder}
\author{T.~Dittrich}
\address{ Depto.\ de F\'\inodot sica, Universidad
de los Andes, A.~A.~4976, Santaf{\'e} de Bogot\'a.}
\author{ B.~Mehlig}
\address{Theoretical Physics, University of Oxford, Oxford, UK}
\author{ H.~Schanz}
\address{Max-Planck-Institut f\"ur Str\"omungsforschung,
Bunsenstr. 10, 37073 G\"ottingen}
\author{ Uzy Smilansky}
\address{{Department of Physics of Complex Systems,} \\
{The Weizmann Institute of Science, Rehovot 76100, Israel}}
\author{ P{\'e}ter Pollner and G\'abor Vattay}
\address{E{\"o}tv{\"o}s University, Department of Physics of Complex Systems \\
H-1518 Budapest, Pf. 32, Hungary}
\date{\today }
\maketitle

\begin{abstract}
We study the spectral statistics for extended yet finite quasi 1-d systems
which undergo a  transition from periodicity to disorder. 
In particular we
compute  the spectral two-point form factor, and the resulting expression
depends on the degree of disorder. It interpolates smoothly between
the two extreme limits -- the approach to Poissonian statistics in the 
(weakly) disordered case, 
and the universal expressions derived in \cite {altsh} and \cite{pa12}
for the periodic case. The theoretical results
agree very well  with the spectral statistics 
obtained numerically for chains of chaotic billiards and graphs. 
  \\
\end{abstract}

}

\section {Introduction \label{INTRODUCTION}}

 The spectrum of an unbounded periodic system is arranged in {\it
continuous} bands
and the corresponding eigenfunctions are {\it extended} (unnormalisable).
When
sufficient disorder is introduced, the system is Anderson localised - the
spectrum is  {\it point like} and the eigenfunctions are {\it localised}
 (normalisable).
The transition from a continuous to a point  spectral measure is a drastic
effect,
which might have been used to characterize the transition. However, this
approach
is of a limited value, since in practice one always deals
with {\em finite} systems, where the spectral measure is point-like both in the
periodic and in the disordered situations. In finite systems, the mean spectral
density is independent of the degree of disorder. Therefore, 
for finite systems, the effect of disorder on the energy spectrum can be
discerned only in the {\it spectral correlations}. 
Indeed, this approach to the characterization of the Anderson transition in
three dimensional systems was used \cite {shklov}, and the
spectral measures were shown to undergo an abrupt change when the critical
level of disorder is reached.  In the  present
paper we study the spectral statistics for finite 
quasi 1-d systems  which undergo a transition from periodicity  to  disorder.
(Quasi 1-d disordered systems of finite length can be either
``metallic'' or ``insulating'' depending on whether the localization length is 
larger or smaller than the system length. We shall consider only the first case,
and the strength of the disorder will be  restricted accordingly, to the 
range of values which is sometimes called weak disorder). We shall focus
our attention to the spectral two-point form factor, and show  
  that it depends very sensitively on the degree of
disorder, and derive a universal expression, which interpolates continuously
between the periodic and the disordered yet metallic limits.  

The spectral form factor is the main object of our discussion, and
it is defined in the following way. The spectrum is unfolded
 by  introducing the dimensionless energy $\epsilon$, through the 
relation  ${\rm d} \epsilon=\dos (E)\, {\rm d}E$ where $\dos(E)$ is
the mean spectral density.  
The corresponding dimensionless time $\tau$
measures time in units of the Heisenberg time $t_{\rm
H}=2\pi\hbar\dos$. We consider a finite spectral interval
of length $\Delta\epsilon$ centered at $\epsilon_{\rm c}$, and denote its
 characteristic function by  
$\chi(\epsilon-\epsilon_{\rm c})$. 
 Since the mean spectral density of the unfolded spectrum is unity,
the number of states in the interval $\Delta\epsilon $ is  ${\cal N}  =
\Delta\epsilon$. This energy interval should be sufficiently large so that
 ${\cal N} \gg 1$, and sufficiently small so that the mean level density and
 the classical dynamics do not change much as the energy is scanned across it.
The oscillatory part of the spectral density in this interval is 
\begin{equation}
\widetilde d(\epsilon)=\chi(\epsilon-\epsilon_{\rm c})\left[
\sum_{q}\delta(\epsilon-\epsilon_{q})-1\right ].
{\label{eq:dens1}}
\end{equation} 
The Fourier transform  of this function is
\begin{eqnarray}\label{d_tau}
d_c(\tau) 
&=&\int  
\e^{-2\pi\i\epsilon\tau}\widetilde d(\epsilon) \  {\rm d}\epsilon  \nonumber\\
&=&\sum_{q}\chi (\epsilon_{q}-\epsilon_{\rm c})\,
\e^{-2\pi\i\epsilon_q\tau}
-\delta_{\Delta\tau}(\tau) \ .
\end{eqnarray}
The Fourier transform of
the normalized characteristic function is denoted
by $\delta_{\Delta\tau}(\tau)$ and its width  is $\Delta\tau\sim
1/\Delta\epsilon$. The form factor is expressed as
\begin{equation}\label{k}
K(\tau)={1\over {\cal N }}\langle|d_c(\tau)|^2\rangle_c.
\end{equation}
We use  $\langle \cdot \rangle_c$  to denote the spectral average, which is
 taken over the non overlapping  energy
intervals located about a set of  $\epsilon_c$ values. One can also perform
the averaging over any free parameter of the system or over  disorder when
it is introduced.
It can be easily shown that (\ref {k}) is nothing but the
Fourier transform of the spectral two-point  correlation density
\cite{Dit96}.   For a discrete
spectrum  the normalization in (\ref{k}) is such that the
form factor approaches a constant $\gamma$ as $\tau\to\infty$ where
$\gamma$ is the
mean spectral degeneracy.

 The expressions for the spectral form factors in the extreme situations of
 exact periodicity and weak disorder are known. In the latter case,
when the length of the system does not exceed the localization length,
and assuming that the Heisenberg time is shorter than the Thouless time 
the spectral statistics  takes the form \cite {ds}
 \begin {equation}
K (\tau)= \left \{
 \begin{array}{ll} g_{\rm T} \sqrt{\tau/ 2 c  }  &{\rm for} \ \ \ \tau < 1  \\
               1      &{\rm for} \ \ \ \tau > 1 \ \ . 
 \end{array}
          \right.
\label {eq:k(t)metalic}
\end{equation}
 The factor $g_T$ can take the values $1$
 or $2$ depending on whether time reversal invariance is 
respected or violated, and $c$ is the conductivity of the chain. 
The spectral form factor for periodic systems was recently derived
using both field-theoretical methods \cite {altsh} and the
semiclassical approximation  \cite{pa12}. Since the latter theory is the basis
for the approach developed in the present paper, we shall describe it briefly
to introduce the concepts and the notations which will be used in the sequel.

 We consider a chain of $N$ identical chaotic unit  cells of length $a=1$,
with periodic boundary conditions, such that the full system shows a
discrete translation invariance (Fig.~\ref{fig:chain}(a)). (Alternatively,
we could discuss a disordered ring configuration which is threaded by
an Aharonov-Bohm flux line. This is the system analyzed in \cite {altsh}).
In such a system, the classical evolution within a unit cell becomes ergodic
after a short time, and one can approximate the classical evolution
in the entire chain by diffusive evolution. We shall denote the diffusion 
constant by $D$. The time it takes the diffusive evolution to cover the 
phase-space uniformly is the Thouless time.  

\begin{figure}[htb]
\centerline{
\psfig{figure=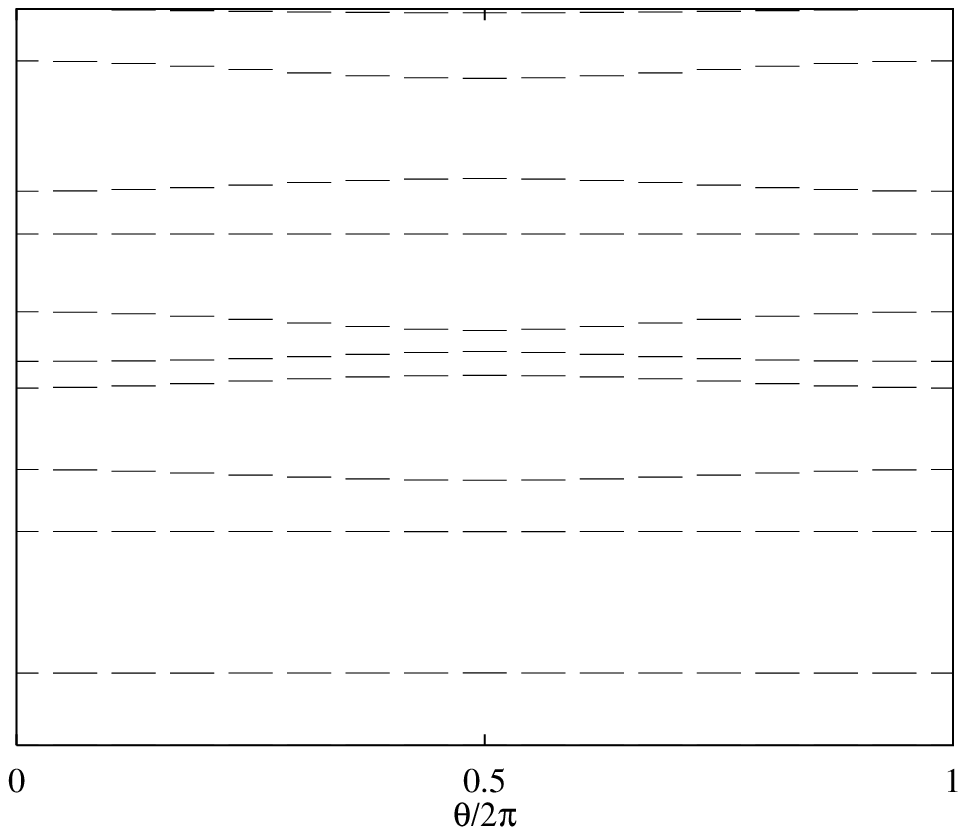,width=4cm}
\psfig{figure=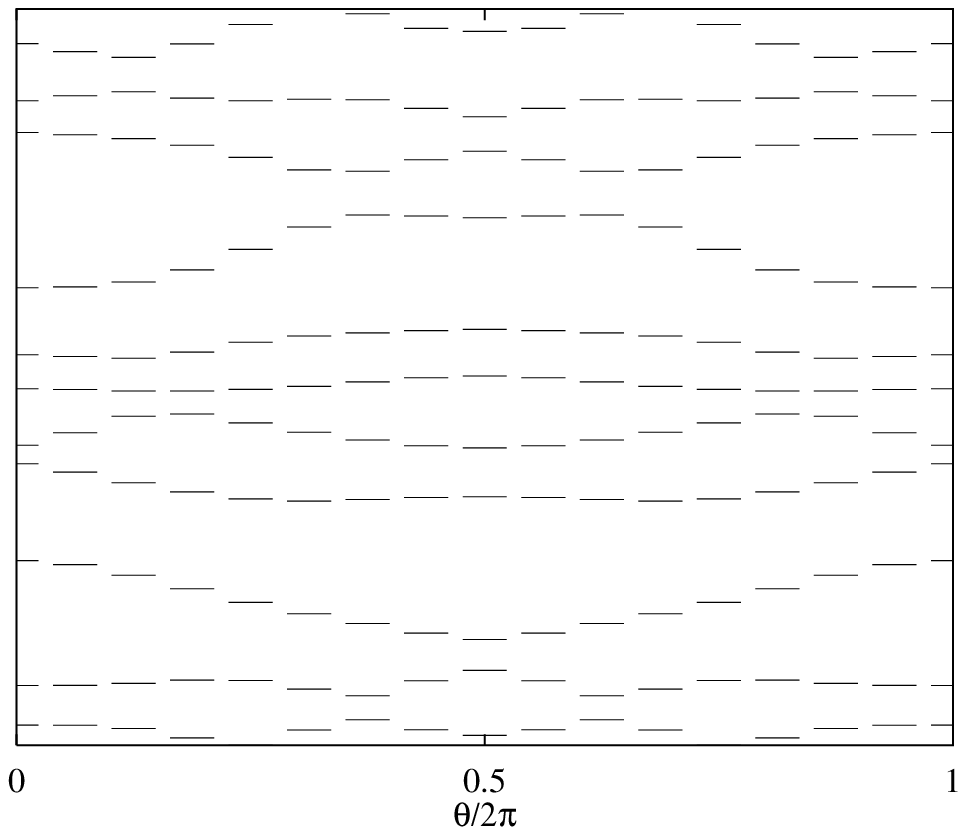,width=4cm}
\psfig{figure=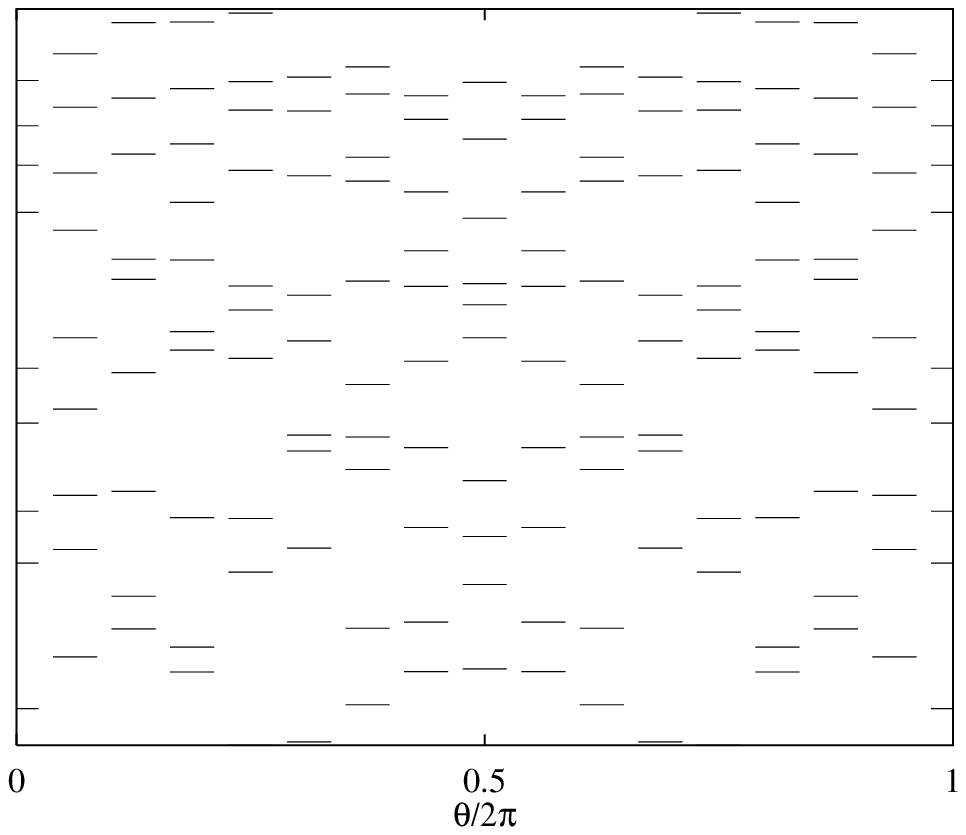,width=4cm}
}
\caption{\label{spag} Typical discretized band spectra of a periodic chain
with $N=16$ unit cells.  The energy levels are shown as a function of the Bloch
phase $\theta_n$ for 10 bands in the case of (a) low, (b) intermediate and (c)
high conductance.}
\end{figure}

Due to translation invariance, the quantum spectrum consists of discretised
energy bands whose width depends on the (dimensionless) conductivity per 
unit cell. It is defined as  $c_1= 2\pi \hbar \langle d_1 \rangle D /a^2$,
where $\langle d_1 \rangle $ is the mean level density {\it  per  unit cell}. 
A few examples of typical bands are shown in 
Fig.~\ref{spag}. One can see that 
 for low $c_1$, the bands are flat and show little
structure. For high values, the bandwidth is of the order of the inter-band
spacing, and the bands can hardly be recognized if the discretisation is too
coarse. 

If the system  under discussion  is invariant under an anti-unitary
symmetry (such as e.g., time-reversal) the bands  are symmetric about  the
center and the edges of the Brillouin zone, and the levels are doubly
degenerate ($\gamma =2$). The reflection symmetry and the degeneracies
 are broken  if the symmetry is lifted, and in this case  $\gamma =1$.

The quantum spectrum is characterized by two energy
scales, the mean {\it intra-band} spacing  and the mean {\it inter-band}
spacing. The ratio between them is at least $N$, the number of  unit cells.
We are interested here in the large $N$ limit, and therefore these energy
scales are very well separated. Since $\dos \approx \langle d_1\rangle  N$, 
the spectral correlations which pertain to the {\it inter-band} scale affect 
the
behavior of the form factor in the range $0 < \tau < 1/N$. The correlations
between levels in the same band leave their mark on
 $K(\tau)$  in the domain $1/N< \tau < 1 $. The fact that the
spectrum is composed of discrete (possibly degenerate) energy levels is
expressed
in the spectral form factor in the domain  $1 < \tau$, where the form factor
approaches the constant value $\gamma$.

 We used different approximations to express the form factor in the three
domains mentioned above \cite {pa12}.
\begin {itemize}
\item  $ 0< \tau <1/N \ $: Here one starts from the semiclassical trace
formula
\cite{Gut71}  and employs the ``diagonal approximation'' \cite{Ber85} to write
\begin{equation}
\label{eq:kdiagon}
 K (\tau) \approx g_{\rm T} N \tau P (\tau) \ .
\end{equation}
The factor $N$ is due to the discrete translation symmetry, because of which
any generic periodic orbit is replicated $N$ times in the system.  $g_{\rm
T}$
stands for the classical degeneracy due to time-reversal (or any other
anti-unitary) symmetry and it can
take the values $1$ or $2$. $P (t)$  is the classical probability to
stay in the same unit cell from which the trajectory started,
 after the time $t=\tau t_H$ \cite{ds}. Because phase-space is covered
diffusively,
$ P (t) \approx ({1\over 2\pi  D t})^{1/2}$
and hence,
\begin{equation}
\label{eq:kdomain1}
 K (\tau) \approx g_{\rm T} N \sqrt{N \tau/ 2 c_1 }   \ ,
\end{equation}
where $c_1$ is the dimensionless conductivity per unit cell which was introduced
above.
\item $ 1/N <\tau <1 \ $: As $\tau$ increases, the form factor  provides
information on a finer energy scale. In the vicinity of $\tau=1/N $, the
energy levels within a single band cannot be resolved, hence $K(\tau
\approx 1/N)$
takes a value which is proportional to the apparent degeneracy $N$. Finer
details
of the energy correlation inside the band are manifested for larger values
of $\tau$. To understand the behavior of the spectral form factor, one
writes the levels in the band $\beta$ as 
$\epsilon_{\beta}(q) \ , \ q=1,\ldots,N$, 
and substitutes in (\ref {k}). Neglecting the cross-band correlations one gets
\begin {equation}
K(\tau)= \left \langle \  {1\over N}
\left | \sum _{q=1}^N {\rm e}^{- i 2\pi \epsilon_{\beta}(q)  \tau}
\right |^2 \ \right \rangle _{\beta}
\label {eq:kdomain2i}
\end{equation}
This is the spectral form factor for a band, averaged over all the bands.
The $q$
summation can be performed by the saddle-point (or the uniform) approximation.
The main contribution comes from the vicinity of the band extrema which
correspond
to the energy values where the spectral density is singular. That is, the
prominent features in the form factor are due to the Van Hove singularities.
Denoting by $\partial^2_q \epsilon_{\beta}$ the second derivative of the band
function at its extrema, one  gets
\begin {equation}
K(\tau)=
C \left \langle (\partial^2_q \epsilon_{\beta})^{-1}\right \rangle _{\beta}
 \tau^{-1},
\label {eq:kdomain2}
\end{equation}
 where $C$ is a numerical constant. It was shown in \cite {pa12} that the 
values of the constants which appear in  (\ref {eq:kdomain1}) and in 
(\ref {eq:kdomain2}) are compatible so that the
two expressions match at $\tau=1/N$.

\item $\tau >1 \ $: The time interval is sufficient to resolve the point-like
character of the spectrum. Hence,
\begin {equation}
K(\tau)= \gamma \ .
\label {eq:kdomain3}
\end{equation}

\end{itemize}

In the following sections we shall study how the expressions
( \ref {eq:kdomain1},\ref {eq:kdomain2},\ref {eq:kdomain3} ) make the
transition to
the Poisson form factor $K(\tau)=1$ as disorder is introduced. The
semiclassical
(diagonal) approximation will be the starting point for the discussion of
the transition in the first domain. This will be done in
section  II. To investigate $K(\tau)$ in the second and the third domains, it
 suffices to study a system which has a single band in the periodic limit.
The $N$--site periodic Anderson model is such a system, and it will be
discussed
in section  III. The important observation made in this section is that
the transition  is well described by considering the disorder
perturbatively. The resulting explicit formulae for $K(\tau)$ in the transition
regime, reproduce the numerical data extremely well. The perturbative
treatment also sheds light
on the peculiar mechanism which reduces the value of $K(\tau)$ from $\gamma$
to $1$ in the third domain when the disorder splits up the degeneracies of the
spectrum. We shall compare the results obtained separately
for the three domains with numerical data for billiard and  graph (network)
systems.
This will be done in section IV, where we shall summarize and discuss our
findings.

\section {{\bf Introducing disorder -- the  semiclassical approximation}
\label{GSCTHEO}}

 We shall compute the spectral form factor (\ref{k})
in terms of the Fourier transform of the oscillatory part of the spectral
density.
Using Gutzwiller's trace formula,  $d(\tau)$ can be expressed
semiclassically as a
 sum over the periodic orbits $j$ of the system
\begin{equation}\label{d_tau_sc}
d(\tau)=\sum_{j}\delta_{\Delta\tau}(\tau-\tau_{j})\;
\tau_{j}\,A_{j}\,\e^{\i\,s_{j}}
\end{equation}
with primitive period $\tau_{j}\approx\tau$. $A_{j}$ denotes the weight of the
orbit corresponding to its stability and includes the Maslov phase. $s_j$ is
the action of the orbit in units of $\hbar$. Following the
standard approximation, we neglect  the contribution of repetitions of
primitive
orbits to the sum  (\ref {d_tau_sc}).   The form factor is now given by a
double sum over periodic orbits
\begin{equation}\label{k_scl}
K(\tau)={1\/\Delta\epsilon} \left
\langle\sum_{j,j'}\delta_{\Delta\tau}(\tau-\tau_{j})\,
\delta_{\Delta\tau}(\tau-\tau_{j'})\,A_{j}A^{*}_{j'}\,\e^{\i\,(s_{j}-s_{j'})}
\right \rangle  \,.
\end{equation}
It is well known \cite{Ber85}, that for short time $\tau$ this sum can be
restricted to the diagonal terms $j=j'$. However, when due to a symmetry,
the orbit appears in $g_j$ {\em different} but symmetry-related versions, the
contribution of all the symmetry-conjugated orbits must be added coherently.
In such cases,  (\ref{k_scl}) reduces within the  diagonal
approximation to
\begin{equation}\label{k_diag}
K(\tau)\approx\sum_{j}g_{j}\,\delta_{\Delta\tau}(\tau-\tau_{j})\,
\tau_{j}\,|A_{j}|^{2}\,.
\end{equation}
In the case of an extended, nearly periodic system, the diagonal approximation
is valid up to $\tau=1/N$, the Heisenberg time of the unit cell \cite{Dit96}.

>From very general arguments it is clear, that in a system whose
phase-space decomposes into several equivalent subspaces related by
(unitary as well as anti-unitary) symmetry, the mean degeneracy $g$ is
 just the number of such subspaces.  Thus, if
time-reversal invariance is the only symmetry obeyed,  phase-space points
with opposite momenta are equivalent and consequently phase-space 
is partitioned in $g_{\rm T}=2$ subspaces.
In our problem,  phase-space is invariant under a  symmetry group
containing $N$ elements and therefore $g=N  g_{\rm T}$.
Using the  sum rule for periodic orbits \cite{HOdA84,ds}, the
form factor is finally written  as
\begin{equation}\label{k_p}
K(\tau) \approx   g_{\rm T} N \tau\,P(\tau) \ .
\end{equation}
which we introduced in the previous section 
(\ref {eq:kdiagon}). 
The normalisation
of the staying probability $P(\tau)$ is such that $P(\tau)=\Omega/\omega(\tau)$
 at a time $\tau$,
where the classical flow covers ergodically the part $\omega(\tau)$ of the
total energy-shell volume $\Omega$. In particular, $P(\tau\to\infty)=1$ for an
ergodic system and $P(\tau)=N$ in a system which is composed of $N$
unconnected ergodic cells. A more precise definition can be found in
 \cite{ds,DS91,Dit96}. For the present purpose, we need only the
following property \cite{Dit96}: the return probability
for a system composed of chaotic unit cells is {\em independent} of the
presence or absence of long-range spatial order.
Thus, within the diagonal approximation, the only
effect of the introduction of disorder is the destruction of the
coherence between the contributions of orbits which were related by symmetry
in the original periodic system. This implies that in the diagonal
 approximation for $K(\tau)$, (see (\ref {k_p})) $g= N  g_{\rm T}$, is to be 
replaced by $g=  g_{\rm T}.$

In order to describe the transition from $g=Ng_{\rm T}$ to
$g=g_{\rm T}$ as the spatial symmetry is broken, we go slightly
beyond the diagonal approximation (\ref{k_diag}) in that we retain in
Eq.~(\ref{k_scl}) the off-diagonal contributions from all those orbits
which are degenerate in the symmetric system
\begin{equation}\label{k_gda}
K(\tau)\approx g_{\rm T}\sum_{r}\delta_{\Delta\tau}(\tau-\tau_{r})\,
\tau_{r}\,|A_{r}|^{2}\,
\left|\sum_{j,j'=1}^{N}\e^{\i\,(\Delta s_{r,j}-\Delta s_{r,j'})}\right|^{2}\,.
\end{equation}
$r$ runs now over all groups of symmetry-related orbits, while $j,j'$ label
the $N$ orbits within each group. Possible degeneracies due to time-reversal
are not affected by breaking the spatial symmetry and are thus contained in
the prefactor $g_{\rm T}$. The disorder which breaks the symmetry has been
assumed weak enough such that (i) the orbits within the Heisenberg time of the
unit cell $\tau_{r}<1/N$ are structurally stable, i.~e.  no (short) periodic
orbits appear or disappear due to the disorder, and (ii) the  
disorder does not alter by much the stability amplitudes and the periods within
a group $r$  so that {\em in the prefactor}  $A_{r,j}\approx A_{r}$, 
$\tau_{r,j}\approx\tau_{r}$. The variation of the actions are of the same
order, but they cannot be neglected because they are measured in units of
$\hbar$
and therefore, the resulting changes in phase, $\Delta s_{r,j}$ should be
taken into account.

Comparing (\ref{k_diag}) and (\ref{k_gda}) we see that Eq.~(\ref{k_p})
represents the form factor also in the case of a weakly broken spatial
symmetry, if $g$ is replaced by an effective degeneracy
\begin{eqnarray}\label{ga_tau}
g(\tau,\delta)&=&{g_{\rm T}\/N}
\left\langle\sum_{j,j'=1}^{N}\e^{\i\,(\Delta s_{r,j}-
\Delta s_{r,j'})} \right\rangle_{r}\nonumber\\
&=& {g_{\rm T}\/N}\(N+\left\langle\sum_{j\ne j'}\e^{\i\,(\Delta s_{r,j}-
\Delta s_{r,j'})}\right\rangle_{r}\)
\end{eqnarray}
which depends on the time $\tau$ since the average on the r.~h.~s.\ is over
all groups of periodic orbits $r$ with length $\tau_{r}\approx\tau$. The
dimensionless parameter $\delta$ has been introduced to characterize the
strength of the symmetry-breaking disorder in a way to be specified in
Eq.~(\ref{do_strength}) below.

In order to evaluate Eq.~(\ref{ga_tau}), we need some information about the
distribution of the disorder contributions to the phases of the periodic
orbits $\Delta s_{r,j}$.  We assume, that the correlation length of the
disorder is negligibly small compared to the mean length of an orbit. In this
case $\Delta s_{r,j}$ is a sum of many independent contributions, and the
number of these contributions is proportional to the period of the orbit
$\tau$. Hence, according to the central limit theorem, $\Delta s_{r,j}$ are
independent Gaussian random variables with mean value $\<\Delta s\>=0$ and
variance
\begin{equation}\label{do_strength}
\<\Delta^2 s\>=\delta^2\tau\,,
\end{equation}
where the average is over all orbits of period $\tau$. With these assumptions
we find from (\ref{ga_tau})
\begin{eqnarray}\label{gav}
g(\tau,\delta)&=&{g_{\rm T}\/N}
\(N+N(N-1)\left|\<\e^{i\Delta s}\>\right|^2\)\nonumber\\
&=& g_{\rm T}(1+[N-1]\,\e^{-\delta^2\tau})\,.
\end{eqnarray}
In the first line we have used the fact that the $N\gg 1$
together with the statistical
independence of $\Delta s_{r,j}$ and $\Delta s_{r,j'}$ justified above to
replace the sum over $j,j'$ by its averaged value. In the second line, the 
Gaussian
distribution of $\Delta s$ was employed to give $\langle \e^{\i\Delta
s}\rangle= \e^{-\langle\Delta^2 s\rangle/2}$. It is easy to see, that
Eq.~(\ref{ga_tau}) indeed interpolates between $g=N  g_{\rm T}$
and $g=g_{\rm T}$  as a function of the disorder. Note that the 
parameter which characterizes the  disorder,  $\delta^2$, is multiplied 
by the time $\tau$  over which the disorder acts. Hence, the classification
of the disorder as ``weak'' or ``strong'' depends on the relevant time scale.

In summary,  we get,
\begin{equation}\label{ksemiclass}
K(\tau,\delta)=  g_{\rm T}(1+[N-1]\,\e^{-\delta^2\tau}) \tau\,P(\tau)
= g_{\rm T}(1+[N-1]\,\e^{-\delta^2\tau}) \sqrt{N \tau/2c_1}\ \  ;  
\ \ \tau < 1/N  \ .
\end{equation}
This expression provides the smooth transition from the periodic 
case, via the weakly disordered to the ``metallic'' domain. 
In section IV we shall show that this simple formula reproduces the
form factor in the   transition from periodicity to disorder
very well. We emphasize once
again that the present theory does not describe  
strongly disordered systems where the localization length is shorter than the
system size. Such systems are outside of the scope of the present approach
 which is  based on the ``diagonal'' approximation.

\section {\bf Introducing disorder - perturbation of a model with a single
band \label{GOBTHEO}}

As explained in the introduction, the form factor in the domain
$\tau > 1/N$ is sensitive to the correlations among the
levels which belong to the same band. Therefore, in order to
investigate the form factor in this region, it is sufficient to study a model
with a single band, which is what we do in the present section.
In order to use the results of this section in the general context, we
have to remember that the form factor in realistic systems is obtained
as an average over many bands (see (\ref {eq:kdomain2i})). This will smooth
out several features of the single-band form factor, as will be explained
in the sequel.

The system we consider is a chain of $N$ unit cells of length
$a=1$, with periodic boundary conditions at the end of the chain.
The chaotic scattering process in each cell is represented by
a random potential and the dynamics  is discretised
on a lattice. Choosing convenient units,
the Schr{\"o}dinger equation reads
\begin{equation}
-\left( \phi_{n+1}-2\phi_n+\phi_{n-1}\right)+
V_n\phi_n=E \phi_n,\quad \phi_n=\phi_{L+n},
\label{schrod}
\end{equation}
where $\phi_n$ is the wave function on the $n$th site. The on--site
potentials $V_n$ are uncorrelated, random variables which are
picked out from the same Gaussian distribution function with variance
$\sigma$.
 They obey
\begin{eqnarray}
\langle V_n \rangle &=&0,\quad \langle V_n V_m\rangle
=\delta_{nm}\sigma^2
\qquad n,m=0\ldots N-1,
\end{eqnarray}
The complexity of the scattering process is incorporated by neglecting
the correlations between the  potentials
on different sites.

In the periodic limit  ($\sigma=0$),
the levels are arranged in a discrete band
\begin{equation}
E_q^0=2(1-\cos(2\pi q/N))\;\;\; q=0,...,N-1.\label{ener}
\end{equation}
The level density (compare Eq.~\ref{eq:dens1})
$$d(E)=\sum_{q=0}^{N-1}\delta(E-E_q)$$
exhibits van Hove singularities at the band edges $E=0$ and $E=4$.
This is  a direct consequence of the periodicity of the system.

For $\sigma\neq 0$ the singularity is smoothed out and for
large values of $\sigma$ the level density becomes uniform between
the upper and lower ends of the spectrum.
This is the typical behavior we expect in generic one-dimensional
disordered systems.
>From here on we consider the periodic case as the limit of
the disordered system when $\sigma\rightarrow 0$.
Accordingly, levels can be unfolded with the constant density.
Since we consider here only weak disorder the mean level density
is taken as
\begin{equation}
\dos \approx N/4.
\end{equation}

\begin{figure}
\centerline{\psfig{figure=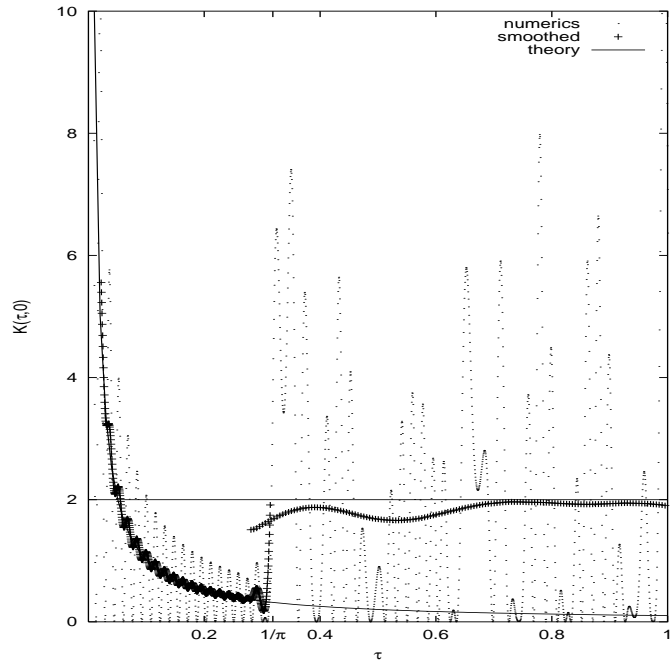,width=9cm,height=9cm}}
\caption{
\label{fig:vat1} 
The form factor (see Eq. 23) for the unperturbed system with lenght $N=64$ 
(dots), the
corresponding smoothed data (points) and the approximation Eq. 26 and 27 (lines).
For $\tau<1/\pi$ the first
term in the Bessel function expansion dominates the form factor.
After averaged over $1/\pi N$ in $\tau$, the smoothed data fit to the
theoretical function (\ref{avek}).
For $\tau>1/\pi$ the soothed form factor converges to $\gamma\approx 2$.
Here the average is taken over a window of $1/\pi$ in $\tau$.
}
\end{figure}

The spectral form factor  was defined in  (\ref {k}).
In the periodic case the energies are given by (\ref{ener}).
The spectral form factor is
\begin{equation}
K (\tau,\sigma=0)=\frac{1}{N}\left|\sum_{q=0}^{N-1} \exp(-i\pi
E_q^0\tau N/2
)\right|^2,\label{kappa0}
\end{equation}
The second argument of the form factor denotes the strength of the
disorder, and in the present periodic case it is $0$.
The expression (\ref{kappa0}) can be rewritten by expanding the
exponential into a Bessel series:
\begin{equation}
e^{iN \tau\pi\cos{\frac{2\pi}{N}q} }=\sum_{k=-\infty}^{\infty}i^k
e^{iq\frac{2\pi}{N}k}\cdot J_k(\pi N\tau)
\end{equation}
Exchanging the order of the summation over $q$ and $k$ yields
\begin{equation}
K (\tau,0)=
N\left|J_0(\pi\tau N)+ 2\sum_{n=1}^{\infty}i^{nN}J_{nN}(\pi\tau N)
\right|^2.
\label{kappbess}
\end{equation}
The function $K(\tau,0)$ which is shown in Fig.~\ref{fig:vat1} 
displays different
features in the three domains of $\tau$.  In the domain $0 \le \tau \le 1/N$
the first term in (\ref{kappbess}) is dominant and $J_0(\pi\tau N)\approx 1 $.
Hence, the form factor assumes the constant value $N$, and does not show
any structure at all because we are dealing with a model with a single
band.
At $\tau >1/N$, the form factor is a highly irregular function. It
fluctuates more rapidly with increasing number of sites $N$. However,
in order to compare the present theory with results which are
derived for realistic  systems, one should remember that in the
latter case, the form factor is averaged over many bands which
differ in their widths and structure. Such averaging can be effectively
achieved by smoothing $K(\tau,0)$ over a small $\tau$ window.

The smoothed form factor $\langle K (\tau,0) \rangle_{\tau}$ is shown in 
Fig.~\ref{fig:vat1}.
In the range  $1/N<\tau <1/\pi $   (\ref{kappbess}) is dominated by the
Bessel function with zero index. 
The average behavior for large $N$
and  $\tau<1/\pi$  can be approximated as
\begin{equation}
\langle K(\tau,0)\rangle_{\tau} \approx
N\langle \left|J_0( \pi\tau N\right|^2\rangle_{\tau} \approx
  \frac{1}{\pi^2\tau},
\label{avek}
\end{equation}
where we used the asymptotic form of the Bessel function
$$ J_\nu(z)\approx \sqrt{\frac{2}{z\pi}}\cos(z-\nu\pi/2-\pi/4)$$
and the average $\langle \cos^2\rangle_{\tau}=1/2$.

In the third domain,  $\tau > 1/\pi$, the window-averaged
function  $\langle K(\tau,0) \rangle_{\tau}$ converges to a
constant value.
This constant is $\gamma$, the average degeneracy of the levels, and it
approximately equals $2$ since most levels are doubly degenerate
(except $E_0=0$ and also $E_{N/2}=4$ if $N$ is even).
In this range of  $\tau$ values,  all Bessel
functions contribute. Resumming the asymptotic forms of the
Bessel functions we get
\begin{equation}
\langle K(\tau,0) \rangle_{\tau} \approx 2
\label{kappasymp}
\end{equation}

We conclude from this discussion that the one-band model, after
proper averaging, reproduces the expected features of $K(\tau,0)$ in
the relevant range  $\tau > 1/N$.

Introducing disorder, the form factor is
given by
\begin{equation}
K  (\tau,\sigma )=
\frac{1}{N}\left \langle \left|\sum_{q=0}^{N-1}
\exp(-i \pi E_q^{\sigma} \tau N/2)\right|^2\right  \rangle_{\sigma }
\label{kappaav}
\end{equation}
where $\langle \ldots \rangle_{\sigma}$ represents the average over 
the disorder and $E_q^{\sigma}$ are the eigenenergies
of the Eq.\  (\ref{schrod}) with $\sigma \ll 1$.

In the case of  weak disorder,
we can use degenerate perturbation theory to calculate how doubly
degenerate energy levels are split. In first order the eigenenergies 
are given by
\begin{eqnarray}
E_{\pm q}^{\sigma}
&\approx & E_{\pm q}^0\pm \frac{1}{N}
\left|\sum_{n=0}^{N-1}V_n \exp(\pm i2\frac{2\pi}{N}qn)\right|
\quad \mbox{for} \quad q\ne 0,N/2
\label{pert1}\\
&\approx &E_q^0 \quad \mbox{for} \quad q=0
\mbox{ (and q=N/2 if N is even)}\label{pert2}.
\end{eqnarray}
Here we ignored a $q$--independent constant, since it does not affect the
form factor.
The main effect of the perturbation is that it
breaks the degeneracy of the energy levels which are symmetrically placed
about the center of the band. The change of the mean level spacing
is small, and can be neglected to leading order.

Substituting the perturbed energy levels into (\ref{kappaav}) and leaving out
the unimportant
$q=0$ (and $q=N/2$ if $N$ is even) levels
we have
\begin{eqnarray}
K(\tau,\sigma)&\approx&\frac{1}{N}\left|
\sum_{q=1}^{\frac{N-1}{2}}2 e^{- \pi iE_q^0\tau N/2}
\cos\left(\frac{1}{2}\left|\sum_{n=0}^{N-1} V_n
e^{\left(i2 \frac{2\pi}{N}qn\right)}\right| \pi\tau  \right)\right|^2 \ .
\end{eqnarray}
The disorder averaging can be performed analytically, as  described in
the appendix.
For large $N$ it leads to
\begin{equation}
  K(\tau,\sigma)=
1+A(2\alpha)+A^2(\alpha)(K(\tau,0)-2),
\label{eq:veg2}
\end{equation}
where the universal function $A(\alpha)$ is defined in the appendix.
A new combination of the variables involving the disorder strength
shows up in this expression
\begin{equation}
\alpha=\frac{\pi\tau \sigma\sqrt{N}}{2},
\label{eq:taualpha}
\end{equation}
governing the properties of the transition from the periodic to 
the disordered case.
 For large $\tau $ values the form factor converges to $1$, since the
perturbation breaks the degeneracy of the levels of the periodic system.
Please note that--as in the semiclassical result Eq.~(\ref{gav})-- the
deviation from the periodic form factor is governed by a dimensionless
parameter containg the product of disorder strength and time.

\begin{figure}
\centerline{\psfig{figure=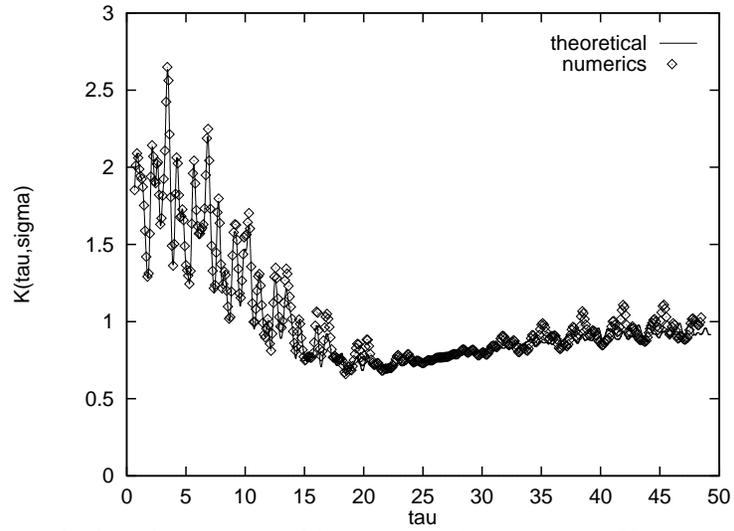,width=10cm}}
\caption{\label{fig:vat4} 
Comparison between the form factor averaged for $1000$ samples
($\sigma=0.008, N=32$) and our perturbative expression. 
Both of them are
smoothed over a $0.5$ window in $\tau$.
Points are from the
numerical calculation, the solid line is the perturbativ formula
of Eq.~(\protect\ref{eq:veg2})
}
\end{figure}

Approximating $K (\tau,0)$ by  its average (\ref{avek}) yields
\begin{equation}
 \langle K(\tau,\sigma)\,\rangle_{\tau } 
\approx
\left \{
 \begin{array}{ll}
(1-e^{-\alpha ^2})^2+ {e^{-\alpha ^2}\over \pi^2\tau } & \mbox{for\ \ } \tau <
1/\pi  \\
  1+A(2\alpha ) &  \mbox{for\ \ } \tau >1/\pi
\end{array}
 \right.  \ .
\label{eq:kappert}
\end{equation}
The $\tau< 1/\pi$ part  describes how the band 
structure is destroyed while the $\tau>1/\pi$ part describes
how the double degeneracy of levels is resolved (see Fig.\ref{fig:vat4}).

We can interpret the result for $\tau>1/\pi$ in terms of the distribution
$p_\sigma(s)$ of splittings of levels $s=E_q^\sigma-E_{-q}^\sigma$ which are
degenerate in the periodic case. For large $\tau$ the form factor is 
the Fourier transform of this distribution:
\begin{equation}
K(\tau,\sigma)=1+\int_0^\infty ds p_\sigma(s)\cos(s\tau \pi N/2).
\end{equation}  
Using the derived  expression (\ref{eq55})  for $A(\alpha)$, we can conclude
 that the splitting distribution has the form
\begin{equation}
p_\sigma(s)= \delta \cdot p_W(s\delta),
\end{equation}
where $p_W(s)=\pi s e^{-\pi s^2 /4}/2$ is the Wigner surmise and 
$\delta=\sigma\sqrt{N}$ is the mean splitting of levels.
The Wigner surmise is known to be the exact distribution of the difference
of the two eigenvalues of a $2\times 2$ GOE random matrices. 
We can conclude that in the present
case,  the ensemble of $2\times2$ matrix describing the splitting of 
levels in the first-order degenerate perturbation theory,
reproduces the spacing distribution of the corresponding GOE
 with mean level spacing $\delta$. 

To check the applicability of the leading order perturbation theory,
we computed the form factor numerically and compared with the analytical
result.
The parameter range was  $\sigma=0.002\ldots 0.256$ and $N=32\ldots 256$.
The numerical results has been averaged for 1000 different disorder
realizations. In Fig.~{\ref{fig:vat4}} we compare formula (\ref{eq:veg2}) and the
simulations. We have found surprisingly good agreement in the whole range of
$\tau$. The  fact that $K(\tau)$ displays a minimum where its value
is less than $1$, and that it approaches $1$ asymptotically from {\em below},
is a direct consequence of the Wigner distribution of level splittings.
In the next section we shall show that this formula  applies very
well also in the case of a multi-banded spectrum, indicating that the
splitting distribution follows the Wigner distribution in more complicated
situations too.

\section {\bf Comparison with numerical results and Discussion
\label{GNUMERICS}}

\subsection{A chain of chaotic billiards}
\begin{figure}[hbt]
\centerline{\psfig{figure=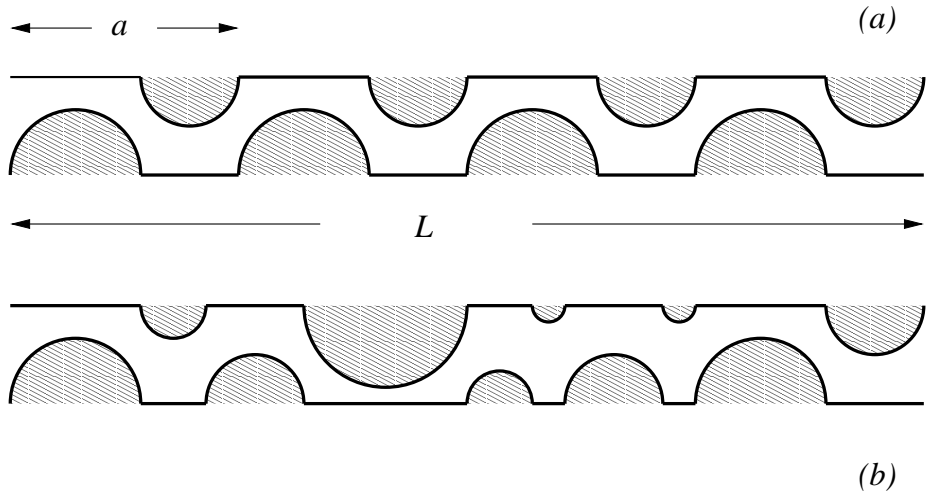,height=8cm}}
\caption{\label{fig:chain} Periodic
(a) and aperiodic (b) chains of
chaotic billiards. The chain length
is denoted by $L$, $a$ is the size
of an individual billiard. Thus
$N = L/a$ is the number
of units in the chain.}
\end{figure}

 The first class of systems  which  were investigated numerically  are
 chains of chaotic billiards,  
(see Fig. \ref{fig:chain})  which can be arranged in a periodic 
(Fig. \ref{fig:chain}(a))  or a disordered (Fig. \ref{fig:chain}(b)) 
fashion. We denote the size of an individual billiard (i.e., the
unit cell in the periodic case) by $a$
and the chain length by $L$ (and $ N =L/a \gg 1)$. 
In the following we discuss weakly disordered chains
and assume that the conductance of the chain  
$c = N c_1   \stackrel{>}{\sim} 1$. On time scales larger than 
the classical ergodic time for a single cell,  the classical
dynamics in the chain of billiards is diffusive,
characterized by the diffusion constant $D$. 
In the diffusive regime, the classical
dynamics of the system, and hence the diffusion constant,
are, to a good approximation, independent of the strength of the disorder. 
The correlations in the quantum spectrum, however,
crucially depend on whether the system
is periodic or not, as discussed above.

In this section we present numerical
results for the form factor $K(\tau)$
in weakly disordered chains. The case of
periodic chains was analyzed in detail in two previous publications
\cite{pa12}. Here we focus on the crossover from 
the periodic case to the weakly disordered (metallic) case, which is predicted
to follow (\ref{ksemiclass}) as a function of the disorder parameter
 $\delta $. Due to time-reversal invariance of the billiard chain,
  $g_{\rm T} = 2$.

\begin{figure}[hbt]
\centerline{\psfig{figure=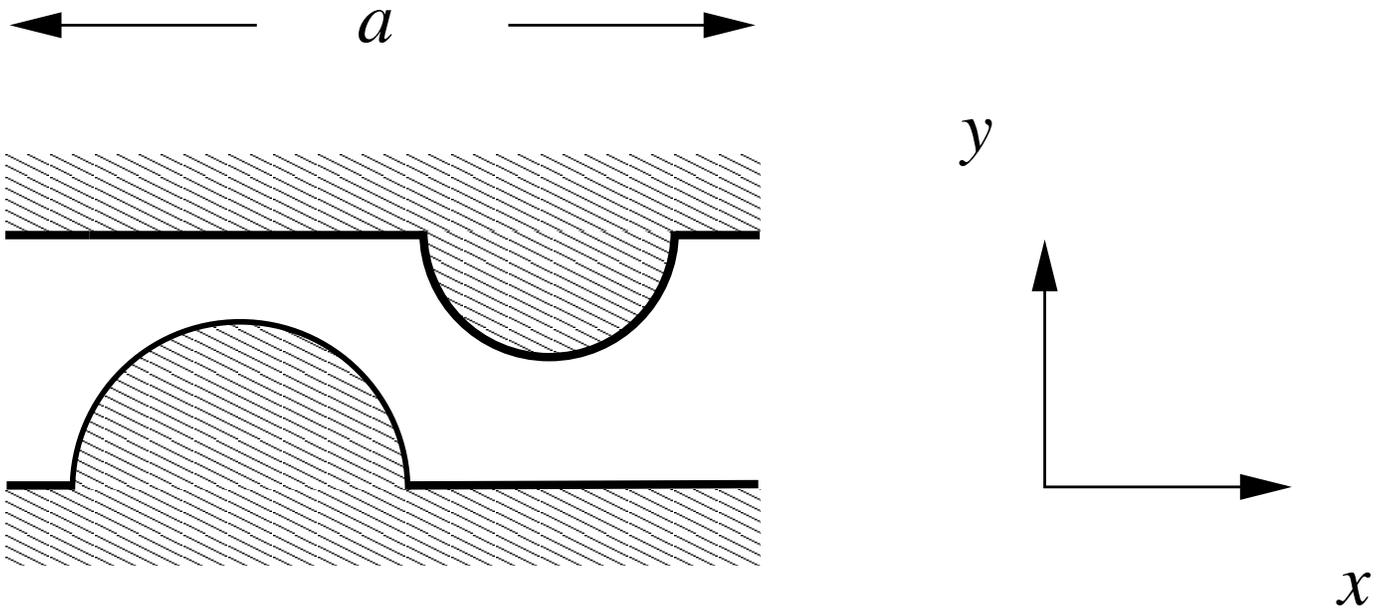,height=8cm}}
\caption{\label{fig:uc} 
Unit cell of the chain
of chaotic billiards.}
\end{figure}

We  have considered a chain composed
of unit cells as shown schematically
in Fig.\ \ref{fig:uc}. The sizes of
the half disks were chosen so that the
contribution of direct trajectories
to the conductance is minimized. Disorder
was introduced by shifting the disks
at random to the right or to the left
by a small amounts $\Delta x$. The dimensionless
variance $\langle( \Delta x)^2\rangle$ is a measure
of the disorder strength. 

Consider a periodic orbit $j$  which hits $M$ 
disks, $m=1,\ldots,M$. Its action $s_j$ (measured in units of $\hbar$) 
is affected by the $M$ shifts $\Delta x_m$
and changes by an amount $\Delta s_j$,
It is plausible that 
$\langle \Delta^2 s_j\rangle 
\propto \langle (k  \Delta x)^2\rangle\, \tau$
and thus 
\begin{equation}
\label{eq:guess}
\delta^2 = C_d  \langle(k  \Delta x)^2\rangle\,.
\end{equation}
We have used (\ref{eq:guess}) to estimate
the quantity $\delta^2$ in  (\ref{ksemiclass}).
The constant of proportionality $C_d$ in (\ref{eq:guess}) remains
undetermined, it depends on the geometry
of the system and on $k $.

We have performed quantum-mechanical calculations for systems composed of 
 $N=16$ unit cells, with disorder parameters covering the domain 
of applicability of (\ref{ksemiclass}).
The quantum-mechanical wave functions satisfy the Helmholtz
equation augmented with Dirichlet boundary conditions
on the channel walls and periodic
boundary conditions along the chain.
The quantum spectrum of this system can be determined
using the method described  in \cite{pa12}. 
In this way we have obtained the quantum
spectra for several realizations of
disorder, as well as for the periodic chain.

\begin{figure}
\centerline{\psfig{figure=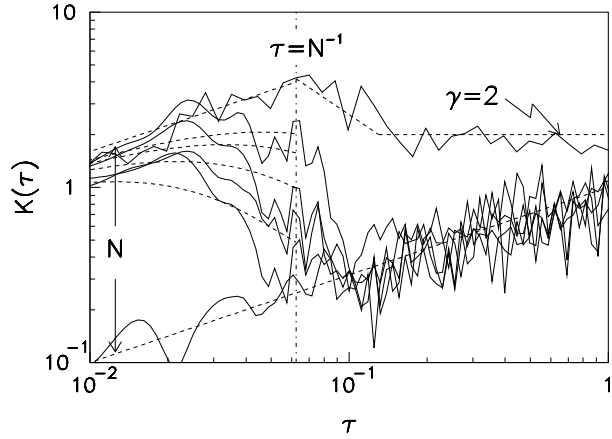,width=10cm}}
\caption{\label{fig:kt} 
$K(\tau)$ for chains
of chaotic billiards and
different strengths of disorder.
}
\end{figure}

Fig. \ref{fig:kt} summarizes the results
of our numerical calculations.  It shows
$K(\tau,\delta)$ as a function of $\tau$,
in the periodic case, for weakly broken
periodicity (four different disorder strengths)
and for weak disorder. The conductivity per unit cell is
independent of the disorder, and its numerical value  was determined 
from a simulation of the classical dynamics 
of the system.  A fit to the diffusion propagator at times
larger than the ergodicity time  allows one to determine
$D$ from which  $c_1 \simeq 33 $ emerges. The calculations 
were conducted for 6 values of the disorder strengths 
$\Delta^2 \equiv \langle k^2 (\Delta x)^2 \rangle = 
0.0, \  6.87\times 10^{-4},\
9.55\times 10^{-4},\ 1.49\times 10^{-3},\ 2.56\times 10^{-3} {\rm and}\  
4.38\times 10^{-3}  $.
In all cases, we have calculated $K(\tau)$ from 1500 eigenvalues,
 in the domain of $k$ values which support $28$ open transverse channels.

The semiclassical theory for the periodic case reproduces the 
numerical results uniformly well over the three ranges of $\tau$ values.
The semiclassical theory matches very well   
the numerical results for the disordered systems in the domain $\tau <1/N$.
However, the numerical results in the domain $1/N < \tau < 1$ are not 
sufficiently smooth to allow a meaningful comparison with the theory
developed in section (\ref{GOBTHEO}). In this domain, the spectrum
is afflicted by frequent near--degeneracies which make the calculation 
 rather costly in terms of computer resources. This problem
is circumvented in the  periodic case, where
the translational invariance is used to facilitate the calculations.
For larger values of the disorder, the degeneracy disappears, but, the
effect we are interested in disappears, too.  
 In the next subsection we discuss a different model
exhibiting a transition from periodicity
to weak disorder, where a quantitative
comparison in the transition regime is possible.

\subsection{A chain  of quantum graphs}
In this section we investigate a second model system---quantized graphs which
were recently shown to provide an excellent example for a quantum chaotic
system \cite{KS97}. The graphs are defined by $v=1,\dots,V$ vertices and
$b=1,\dots,B$ bonds with lengths $L_b$ connecting them.  The wave function on
a graph is a $B$-component function $\left ( \psi_1(x_1),\cdots , \psi_B(x_B)
\right ) ^T$. Each component satisfies the Schr{\"o}dinger equation
($\hbar=2m=1$)
\begin{equation}\label{se}
\({{\rm d^2}\/{\rm d}x_b^2}+k^2\)\psi_{b}(x_b)=0\,.
\end{equation}
At the vertices, the wave function must satisfy boundary conditions
which impose  continuity and current conservation.
They guarantee that the Schr{\"o}dinger
operator is self adjoint, and its spectrum consists of discrete points.
Implementing the boundary conditions, one derives
a secular equation which provides a convenient means to compute
the spectrum numerically. 
The graph is essentially a one-dimensional system, and
therefore, the mean (wave number) spectral density is constant, proportional
to the total length of the graph.

\begin{figure}
\centerline{\psfig{figure=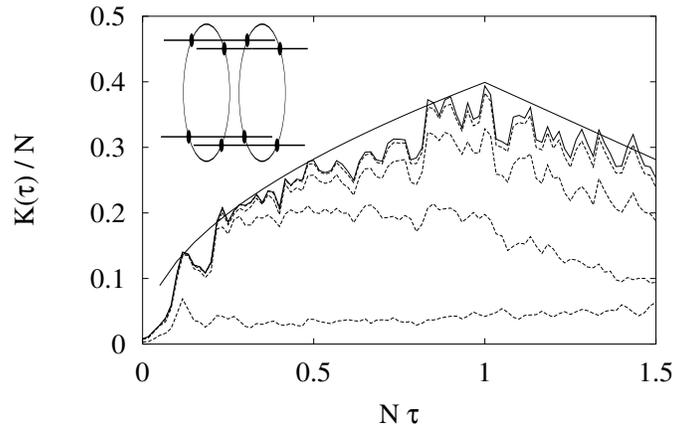,width=9cm}}
\caption{\label{ff_2} The form factor for a quantum graph consisting of N=8
unit cells with $2\times 4$ vertices each. The unit cell is shown in the inset
with each dot corresponding to a vertex.  The form factors were computed using
the lowest 10,000 bands. Each spectral window in Eq.~(\protect\ref{lspec})
contained 30 bands. The disorder strengths were (top to bottom)
$\Delta=0,\,0.02,\,0.05,\,0.1,\,0.5$.  The smooth curve represents the
theoretical prediction without disorder.}
\end{figure}

The graph representing one unit cell was chosen to be the ``cylindrical'' 
network
shown in the inset of Fig.~\ref{ff_2}. The cylinder consists here of $n_{x}=2$
layers with $n_{y}=4$ vertices each.  The unit cell was constructed from more
than one layer in order to remove any residual symmetry. Two bonds lead from
each vertex to the neighboring layers, two more to other vertices in the same
layer. Hence we have for the unit cell $V=n_{x}n_{y}$ and $
B=2n_{x}n_{y}$. The lengths of all bonds are random, but the total length of
the graph was fixed at $L_{\rm H}=2\pi$ such that the mean length of a bond is
$\bar L=\pi(2N\,n_{x}n_{y})^{-1}$ and the mean level spacing with respect to
the wave number $k$ is unity. For this reason it is natural to use, instead of
energy and time, the wave number $k$ and the length $l$ as conjugate variables,
 since
then no unfolding is necessary. In complete analogy to (\ref{k}) we introduce
the spectral form factor via the length spectrum of the oscillating spectral
density $\widetilde d(k)=\sum_{q}\delta(k-k_{q})-1$ using a rectangular window
\begin{equation}\label{ff}
K(\tau)={1\/\Delta k} \<\left|d(\tau)\right|^2\>\,,
\end{equation}
\begin{equation}\label{lspec}
d(\tau)=\int_{k-\Delta k/2}^{k+\Delta k/2}{\rm d}k' \e^{-2\pi\i k\tau}
\widetilde d(k)\,.
\end{equation}
$\tau=l/2\pi$ is simply given by the path length $l$ measured in units of the
Heisenberg length $L_{\rm H}=2\pi$.

The classical analogue for the quantum graph is the random walk of a particle
moving freely along the bonds and scattering at the vertices according to the
quantum transition probabilities \cite{KS97}. 
In the graphs we consider here,
exactly four bonds are attached to each of the vertices. In this case the
transition probability is $1/4$ for all bonds, and the Lyapunov exponent is
$\ln 4$ when time is scaled with the mean time between successive vertex
traversals. The coarse-grained classical evolution is diffusive $\<n_{\rm
w}^2\>=D_{n}n=D_{l}l$, where $n_{\rm w}$ is the distance along the chain
measured in unit cells (i.~e.\ $n_{\rm w}$ is the winding number in the periodic case), $l$
is the length of a trajectory and $n=l/\bar L$. When allowance is made for the
fact that only half of the traversed bonds contribute to the diffusive
transport, the diffusion constant is easily found from the analogy to a random
walk on a 1D-lattice with discrete time: $D_{n}=1/2n_{x}^2$,
$D_{l}=N\,n_{y}/\pi n_{x}$.

The return probability entering (\ref{k_p})  decays as $\tau^{-1/2}$ until
it saturates at $1$ when the diffusion covers the whole chain ergodically.
The number of unit cells $N=8$ was chosen such that this saturation occurs
beyond the Heisenberg time of the unit cell $\tau_{\rm H}^{\rm (uc)}=1/N$ and
is thus not relevant for the form factor. In this case the return probability
is explicitly given by
\begin{equation}\label{}
P(\tau)={N\/2\pi\sqrt{D_{l}\tau}}
\qquad (\tau_{\rm erg}^{\rm (uc)}\le\tau\le{1/N})
\end{equation}
Using $g=2N$ for the mean degeneracy of periodic orbits  we finally obtain
for the form factor
\begin{equation}
K(\tau)=N\,\sqrt{{n_{x}\/\pi n_{y}}N\tau}\qquad (N\tau<1)\,,
\end{equation}
which is shown in Fig.~\ref{ff_2} with a smooth solid curve and has to be
compared to the data obtained numerically without disorder (upper fluctuating
curve). Beyond $N\tau=1$ the smooth curve shows the decay of the form factor
as $1/\tau$. Although the quantitative agreement is not perfect, the theory
reproduces the essential features of the form factor, and in particular the
peak at the Heisenberg time is correctly predicted.

 The disorder was introduced by small changes in
the lengths of all bonds
\begin{equation}\label{delta_l}
L_{b}^{(\Delta)}=L_{b}^{(0)}+\Delta L_{b}
\end{equation}
such that the total length remains constant $\<\Delta L_{b}\>=0$. Here, $b$  
 runs over all the $NB$  bonds of the whole system. The
strength of the disorder is characterized by the dimensionless parameter
\begin{equation}\label{delta}
\Delta^{2}=k^2\<\Delta^2 L_{b}\>\,.
\end{equation}
 As shown in Fig.~\ref{ff_2}, the  peak which characterizes $K(\tau,\delta=0)$
disappears gradually, when the strength $\delta $ of the disorder is increased.
In order to be able to apply the theory developed above, we have to take 
into account a feature which is particular to the graphs system.  
A periodic orbit of length $\tau$ traverses on the average $n(\tau)=\tau
L_{\rm H}/\bar L=4N\tau n_{x}n_{y}$ bonds, which, for sufficiently large
$\tau$, can justify the discussion preceding (\ref{do_strength}) in section
(\ref{GSCTHEO}). We have to bear in mind, however, that in fact not
all of the $n(\tau)$ length variations $\Delta L_{b}$ accumulated in this way
need to be independent, since in general, some of the bonds are traversed
several times and moreover, for time-reversal symmetry the reversed bond
contributes the same variation $\Delta L_{b}=\Delta L_{\bar b}$.  For this
reason we introduce an average bond multiplicity $m(\tau)$ for an orbit of
period $\tau$. Then, the action variation of such an orbit is the sum of
$n(\tau)/m(\tau)$ independent contributions, each with a variance
$m^2(\tau)\Delta^2$. Hence we find for the variance of the sum
\begin{equation}\label{delta_m}
\langle\Delta^2 s\rangle=n(\tau)m(\tau)\Delta^2\,.
\end{equation}

\begin{figure}
\centerline{\psfig{figure=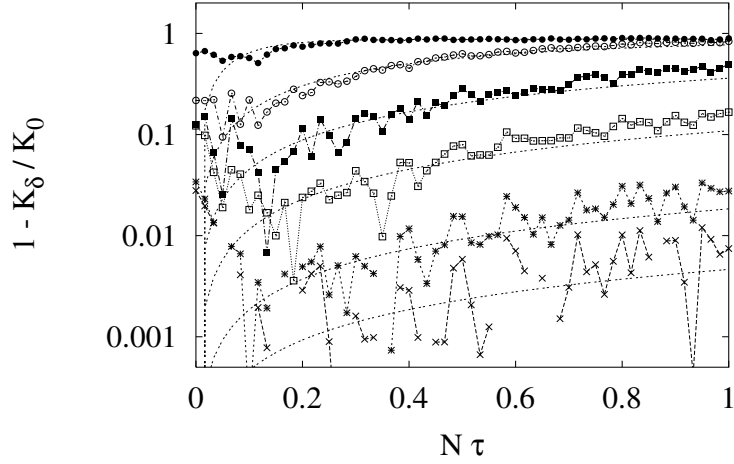,width=10cm}}
\caption{\label{affr_2} Reduction of the form factor due to disorder for
various values of the disorder strength
$\Delta=0.01,\,0.02,\,0.05,\,0.1,\,0.2,\,0.5$ (bottom to top)
compared to the prediction of Eq.~(\protect\ref{gav}).
}
\end{figure}

In order to obtain an estimate for $m(\tau)$ , we assume that a typical orbit
covers ergodically some region of the phase-space such that each bond
is traversed twice on the average (with momentum $\pm 1$) and hence
$m=2$. This is the case, e.~g., at the Heisenberg
time for an isolated unit cell, and---lacking a satisfying theory for
$m(\tau)$---we have no choice but to generalize this special case. 
Comparing
(\ref{delta_m}) with (\ref{do_strength}), we find for the disorder strength
$\delta^2=8N\,n_{x}\,n_{y}\Delta^2$. This is the parameter which we have
chosen in Fig.~\ref{affr_2} in order to compare the numerical data from
Fig.~\ref{ff_2} with the result of section \ref{GSCTHEO}. In order
to better distinguish the curves for small $\Delta$ we plot the quantity
$1-K_{\delta}(\tau)/K_{0}(\tau)$ which is according to (\ref{k_p}) and
(\ref{gav}) given by $(N-1)/N(1-\e^{-\delta^2\tau})$ and find indeed a
reasonable agreement between the theory and the data.

\begin{figure}
\centerline{\psfig{figure=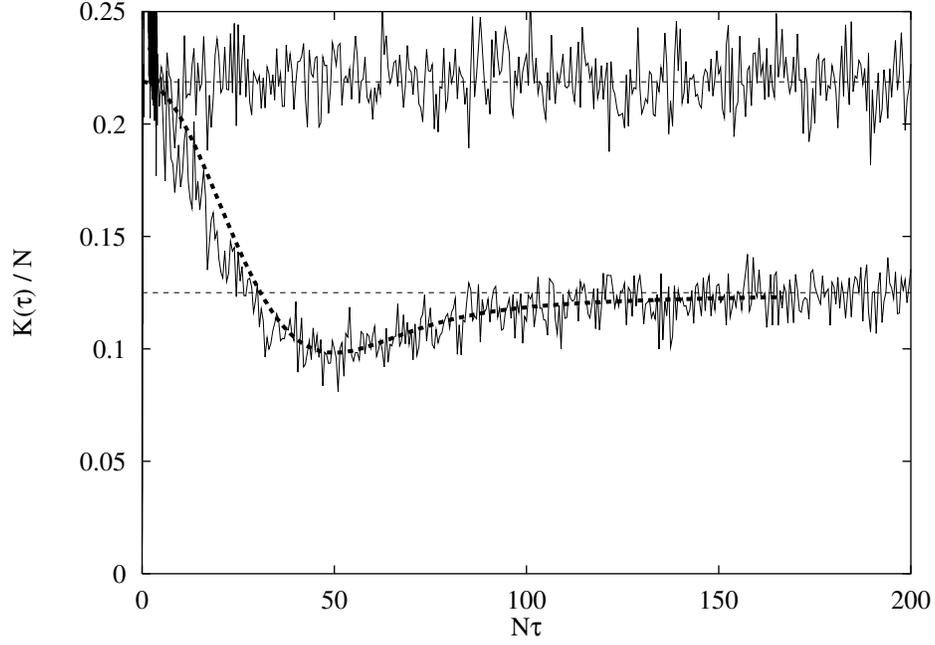}}
\caption{\label{asymp}
Large--$\tau$ behaviour of the form factor for a periodic and a weakly
disordered ($\Delta=0.01$) quantum graph (upper and lower solid lines,
respectively) with $N=8$ unit cells.  The horizontal dashed lines represent
the asymptotic values $K(\tau)=2-2/N$ and $K(\tau)=1$. The heavy dashed lines
show $\langle K(\tau,\sigma)\rangle_{\tau}$ according to 
Eq.~(\ref{eq:finite_n}). The parameter $\sigma$ entering 
Eq.~(\ref{eq:finite_n})
 via Eq.~(\ref{eq:taualpha}) has been
determined by adjusting the the location of the minimum of the function
$A(2\alpha)$ to the minimum observed in the numerical data.  }
\end{figure}

In Fig.~\ref{asymp} we compare the graph data with the perturbative theory for
a single band developed in section III. Since in our numerical
calculations the number of unit cells $N=8$ was not very large, we have to
take into account the fact that for even $N$ two levels in each band---at the
border and in the center of the Brillouin zone---are not degenerate. Only the
remaining $N-2$ levels are described by the perturbative theory of section
(\ref {GOBTHEO}), and consequently Eq.~(\ref{eq:kappert}) is replaced by 
\begin{equation}
\label{eq:finite_n}
K(\tau)=1+{N-2\over N}A(2\alpha)\,,
\end{equation}
such that the asymptotic value in the periodic case is $2-2/N$. Qualitatively,
Eq.~(\ref{eq:finite_n}) predicts that the form factor for the periodic case
has a minimum and beyond that approaches its asymptotic value from below. This
non--trivial behavior is indeed observed in our numerical data. For a
quantitative comparison we had to determine the unknown constant $\sigma$
which relates $\tau$ to $\alpha$ according to Eq.~(\ref{eq:taualpha}).  We have
chosen $\sigma$ such that the position of the minimum in $K(\tau)$ is the same
for theory and numerics. Indeed this leads to a satisfactory agreement of
Eq.~(\ref{eq:finite_n}) with the data, in particular beyond the minimum. It is
reasonable, that this agreement becomes worse for smaller $\tau$, since then
the main assumption behind Eq.~(\ref{eq:finite_n})---the lack of any
correlation between different pairs of nearly degenerate levels---breaks
gradually down.

 Summarizing our findings, we can confidently state that
the numerical results displayed above provide convincing evidence in favor of
the applicability of the simple semiclassical and perturbative approaches. 
This theory grasps the essential features of the transition, and provides 
simple expressions (\ref {ksemiclass},\ref{eq:kappert})
for the form-factor and its dependence on the disorder.
The main drawback of this theory is that it makes use of different 
approximations, depending on whether $\tau$ is larger or smaller 
than the Heisenberg
time $1/N $. In the periodic limit, one could check the applicability
of the theory in the vicinity of the Heisenberg time, by comparing it   
with the field theoretical expression which was
derived for periodic systems which violate time-reversal symmetry.
The field-theoretical treatment \cite {altsh}
 provides an expression which is uniformly valid for the entire
$\tau$ domain. 
The semiclassical theory of \cite {pa12} coincides with
the field-theoretical expression in the separate domains of its validity,
and did quite well even when the two expressions were extrapolated to the
domain $\tau \approx 1/N$. A similar  field-theoretical treatment of the   
the transition from the periodic to the disordered case does not exist yet,
and it is naturally called for.

{\bf Acknowledgments}

We acknowledge support from the Hungarian-Israeli Scientific Exchange program
ISR-9/96 and the Soros Foundation 222-3383/96 for supporting the  visit of PP
at the Weizmann Institute, where this work was initiated.
The support from the Minerva Center for Nonlinear Physics is also 
acknowledged. HS acknowledges the kind hospitality of the WIS during his visit.
GV thanks the Hungarian Ministry of Education and the OTKA T25866/F17166 for 
the financial support.
GV and US thank the Humboldt foundation and PP thanks the KAAD for supporting
their simultaneous visit in Marburg, where some of the results were derived.
They are obliged to Bruno Eckhardt and the Department of Physics
at the Philipps--Universit{\"a}t Marburg for the cordial hospitality.

\section{Appendix \label{APPENDIX}}
\begin{figure}
\centerline{\psfig{figure=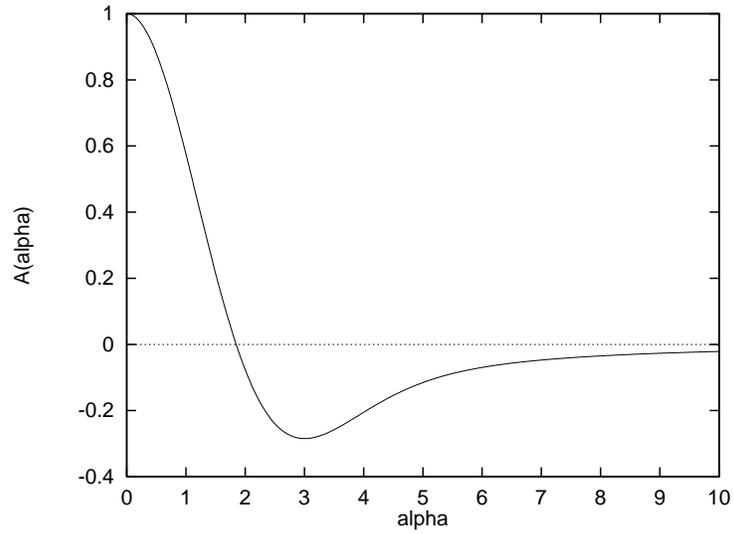,width=10cm}}
\caption{\label{fig:vat5}
The universal function $A(\alpha)$.
}
\end{figure}

For calculating the averaged form factor in the disordered case,
one needs the following quantities:
\begin{equation}
\langle \cos(x)\rangle =\sum_{k=0}^{\infty}\frac{(-1)^k\langle x^{2k}
\rangle}{(2k)!}
\label{cosav}
\end{equation}
and
\begin{equation}
\langle \cos(x)\cos(x^{\prime})\rangle=\sum_{k=0,l=0}^{\infty}
\frac{(-1)^{k+l}\langle x^{2k} x^{\prime 2l}\rangle}{(2k)!(2l)!}
\label{cosav2}
\end{equation}
where
\begin{equation}
x=\frac{1}{N}\left|\sum_{n=0}^{N-1} V_n
\exp\left(i2 \frac{2\pi}{N}qn\right)\right| \pi\tau N/2
\end{equation}
and $x'$ denotes the same, except the $q$ is substituted by $q'$.
Since the Taylor series of the cosine contains only even powers
of its argument, after some simple but tedious calculations using
the properties of Gaussian random distributions, we have the closed
form:
\begin{equation}
\langle x^{2k}\rangle =( \pi\tau  \sigma /2)^{2k} N^{k} k! \ .
\label{evpow1}
\end{equation}
The same calculations also show, that if $q\neq q'$ then the variables
are uncorrelated:
\begin{eqnarray}
\langle x^{2k}x^{\prime 2l}\rangle & = &
( \pi\tau  \sigma /2)^{2k+2l} N^{k+l} k!l! =
\langle x^{2k} \rangle\langle x^{\prime 2l}\rangle
.
\label{evpow2}
\end{eqnarray}
Introduce the parameter
\begin{equation}
\alpha=\frac{ \pi\tau \sqrt{N} \sigma}{2} \ ,
\label{alpdef}
\end{equation}
and using the  property (\ref{evpow2}), we define the function
$A(\alpha)$ which appears in the expression  (\ref {eq:veg2}) for 
 $K (\tau,\sigma)$:
\begin{equation}
\langle \cos(x) \rangle  =  A(\alpha) .
\end{equation}
One can also show that
\begin{eqnarray}
\langle \cos(x)\cos(x')\rangle & = & \langle \cos(x)\rangle \langle
\cos(x')\rangle = A^2(\alpha)\\
\langle \cos(x)\cos(x)\rangle & = & \left\langle \frac{1}{2}+
\frac{1}{2}\cos(2 x)
\right\rangle=
\frac{1}{2}(1+A(2\alpha))
\end{eqnarray}
After substituting (\ref{evpow1}) into (\ref{cosav})
for the $A(\alpha)$ function results\cite{hansen}:
\begin{equation}
A(\alpha)=\sum_{k=0}^{\infty}(-1)^k \alpha^{2k} \frac{k!}{(2k)!}=
1-|\alpha| e^{-\alpha^2/4} {\rm\, Erfi}(|\alpha|/2)=
1+\sqrt{\pi}\frac{i\alpha}{2}e^{-\alpha^2/4}{\rm\, Erf}(i\alpha/2)
\label{eq55}
\end{equation}
where Erfi(x) denotes the error function for imaginary argument,
the Erf(x)  is the commonly used error function.
The behavior of the function (see Fig.~\ref{fig:vat5}) for
 small arguments  is Gaussian:
\begin{equation}
A(\alpha)=e^{\alpha^2/2}(1-\frac{1}{24}\alpha^4+o(\alpha^6))
\end{equation}

\pagebreak

\end{document}